\documentclass[lettersize, journal]{IEEEtran}

\IEEEoverridecommandlockouts

\usepackage{cite}
\usepackage{amsmath,amssymb,amsfonts}
\usepackage{algorithmic}
\usepackage{graphicx}
\usepackage{textcomp}
\usepackage{xcolor}
\usepackage{xr}
\usepackage{multirow}
\usepackage{tabularx}
\usepackage{enumitem}
\usepackage{booktabs}
\usepackage{todonotes}
\usepackage{xurl}
\usepackage{orcidlink}
\usepackage{balance}

\newcommand{\BibTeX}{\textrm{B \kern -.05em \textsc{i \kern -.025em b} \kern -.08em
T \kern -.1667em \lower .7ex \hbox{E} \kern -.125emX}}
\begin{document}

\title{Load-Balancing versus Anycast:\\A First Look at Operational Challenges}

\newcommand{\ie}{\textit{i.e.}}
\newcommand{\etc}{\textit{etc.}}
\newcommand{\eg}{\textit{e.g.}}

\author{
\IEEEauthorblockN{
Remi Hendriks\IEEEauthorrefmark{1}\orcidlink{0009-0008-1578-668X},
Mattijs Jonker\IEEEauthorrefmark{1}\orcidlink{0000-0001-5174-9140},
Roland van Rijswijk-Deij\IEEEauthorrefmark{1}\orcidlink{0000-0002-0249-8776},
Raffaele Sommese\IEEEauthorrefmark{1}\orcidlink{0000-0003-3484-9259}
}
\IEEEauthorblockA{\IEEEauthorrefmark{1}
\textit{University of Twente}, Enschede, The Netherlands \\
\{remi.hendriks, m.jonker, r.m.vanrijswijk,  r.sommese\}@utwente.nl}
}

\newcommand{\manycast}{MAnycast\textsuperscript{2}}

\maketitle
\IEEEpeerreviewmaketitle

\begin{abstract}
Load Balancing (LB) is a routing strategy that increases performance by
distributing traffic over multiple outgoing paths.
In this work, we introduce a novel methodology to detect the influence of LB
on anycast routing,
which can be used by operators to detect networks that experience anycast site flipping,
where traffic from a single client reaches multiple anycast sites.
We use our methodology to measure the effects of LB-behavior on
anycast routing at a global scale, covering both IPv4 and IPv6.
Our results show that LB-induced anycast site flipping is widespread.
The results also show our method can detect LB implementations on the global Internet,
including detection and classification of Points-of-Presence (PoP) and egress selection techniques
deployed by hypergiants, cloud providers, and network operators.
We observe LB-induced site flipping directs distinct flows to different anycast sites with significant latency inflation.
In cases with two paths between an anycast instance and a load-balanced destination,
we observe an average RTT difference of 30\,ms with 8\% of load-balanced destinations seeing RTT differences of over 100\,ms.
Being able to detect these cases can help anycast operators significantly improve their service for affected clients.
\end{abstract}



\section{Introduction} \label{sec:introduction}
Anycast has been widely used in the DNS ecosystem for decades~\cite{dns_anycast}.
The main motivation for deploying such critical Internet infrastructure using anycast is the great resilience it offers by replicating services at multiple discrete locations~\cite{rfc4786}.
However, a pressing concern of anycast is that is only suitable for stateless services (like DNS over UDP) due to anycast routing instability that causes client traffic to reach multiple sites~\cite{site_flipping}.
When a client establishes state with an anycast service, such instability would result in the connection to break as the state is only kept at the initial site.
One cause of such instability is load-balancing (LB) that distributes traffic over multiple available network paths that, in the case of anycast, can result in traffic reaching multiple sites.

Yet, previous work shows that anycast operators provide stateful services over TCP~\cite{tcp_anycast}.
Additionally, we find that DNS operators deploy DNS over TCP, HTTP, and QUIC~\cite{dns_over_x},
suggesting that the resilience that anycast offers can be extended to stateful services.
In this work we take a closer look at anycast routing stability by investigating the effects of load-balancing (LB) on anycast routing.

Load-balancing (LB) is designed to provide redundancy for the Internet by utilizing multiple paths toward a destination.
Additionally, LB avoid under-utilizing certain links by distributing load over available network paths.
An example of this is Equal-Cost Multi-Path (ECMP), where routers that perform load-balancing (\ie, \textit{load-balancers}) split traffic between `equal cost' paths
using a hashing algorithm that is computed over header fields of packets.
Most commonly, the headers used are the five-tuple of flows; destination and source address,
the `next protocol' field, {\ie, TCP, UDP, ICMP, \etc}, and the destination and source port. 
The main motivation for computing hashes per-flow, is that this ensures packets belonging to the same flow traverse the same path
minimizing out-of-order deliveries for the recipient.

To measure the effects of load-balancing on anycast routing we use Verfploeter~\cite{verfploeter},
a methodology that allows for measuring anycast catchments by sending ICMP probes using an anycast source address.
In this work, we extend this methodology to allow for UDP and TCP probing and sending multiple probes to a single target with varied header fields.

Hence, by selectively and iteratively changing header fields using our anycast probing system,
we purposely trigger LB behavior on the Internet.
This allows us to find networks that experience \textit{anycast site flipping}, where traffic reaches multiple anycast sites.
Using our approach, we evaluate the effects of LB on anycast at scale.

Previous work uses traceroute to detect LB
by finding diverging paths when varying header fields~\cite{mda,mdav6}.
However, these approaches suffer from a large probing overhead, making them unsuitable for Internet-wide measurement of LB\@.
Furthermore, traceroute has constraints, \eg, it requires hops to decrement \texttt{TTL} and send \texttt{ICMP Time Exceeded} replies.
These constraints make the approach unsuitable for detecting load-balancing in networks that, \eg, use tunneling, a widely used technique~\cite{mpls}.
Our approach does not depend on traceroute and has a low probing cost,
making it suitable for detecting the effects of load-balancers on anycast routing at Internet scale.
Furthermore, operators can use our methodology to optimize their anycast deployment
and minimize the number of clients that experience \textit{anycast site flipping}.

Our findings show that LB-induced anycast site flipping is prevalent.
We observe clients reach anycast sites in different continents leading to large RTT differences.
We observe hypergiants and tier-1 transport providers, who have sophisticated network infrastructures,
connect poorly with anycast and find that this work's novel methodology can be used to measure cases where this leads to undesirable routing.

Our contributions are a methodology that allows operators to find networks
that experience anycast site flipping, which we validate using Internet wide measurements
that show the performance degradation site flipping causes for anycast,
and make our implementation of the methodology publicly available as part of a larger toolchain~\cite{manycastr_tooling}.

The remainder of this paper is structured as follows. 
First, we provide background information and discuss related work in \S\ref{sec:background_and_rw}.
Then, in \S\ref{sec:methodology} we introduce and detail our methodology.
We discuss our results in \S\ref{sec:results}.
Finally, in \S\ref{sec:conclusion} we conclude and outline future work.


\section{Background \& Related Work}\label{sec:background_and_rw}

\noindent\textbf{Load-Balancing} --
LB is a routing strategy that enables packets to be forwarded toward a
single destination using multiple paths.
This strategy is widely
used in networking to scale up bandwidth on the Internet.
LB can be applied
on a per-packet, per-destination, or per-flow basis.
The per-flow application hashes such that packets belonging to the same flow take the same path,
the per-destination approach implements a coarse-grained approach where packets toward the same destination take the same path,
and the per-packet application load-balances packets individually.
The per-flow approach is most widely
used~\cite{mca}, while the per-packet approach may suffer from out-of-order delivery
and inconsistent MTU issues~\cite{RFC2991}.
The specific hash function depends on the router vendor, specific hardware,
and configuration that can be set by operators.
Table~\ref{tab:vendors} gives an overview of the header fields used by major router vendors~\cite{vendors} when using default configurations.
Juniper hashes using the source address, destination address, and \textit{protocol} fields of the IP header (network layer 3)~\cite{juniper}
and for their \textit{MX series} routers use the entire flow 5-tuple (layer 3 \& 4)~\cite{junipermx}.
Cisco and HPE use the IP address fields (layer 3)~\cite{cisco, hpe}.
Arista, Huawei, and Nokia hash using the flow 5-tuple (layer 3 \& 4)~\cite{arista, huawei, nokia}.
Finally, we observe all vendors have LB options for tunneling protocols such as GRE and MPLS\@.

\begin{table}
\resizebox{\columnwidth}{!}{
\begin{tabular}{|c|r|}
        \hline
        \textbf{Vendor} & \multicolumn{1}{c|}{\textbf{Header fields}} \\
        \hline
        \hline
        	Arista\cite{arista} & Flow 5-tuple (layer 3 \& 4)\\
	\hline
	 Cisco\cite{cisco} & SRC, DST (layer 3)\\
	\hline
	HPE\cite{hpe} & SRC, DST (layer 3)\\
	\hline
	 Huawei\cite{huawei} & Flow 5-tuple (layer 3 \& 4)\\
	\hline
	 Juniper\cite{juniper} & SRC, DST, \textit{protocol} (layer 3)\\
	\hline
 	Juniper (MX series)\cite{junipermx} & Flow 5-tuple (layer 3 \& 4)\\
	\hline
 	Nokia\cite{nokia} & Flow 5-tuple (layer 3 \& 4)\\
	\hline
\end{tabular}
}
\vspace{0.1em}
\caption{The header fields that major router vendors use for their LB hash-function (default configurations).
\vspace{-2em}
}
\label{tab:vendors}
\end{table}

\vspace{0.1em}\noindent\textbf{Anycast Site Flipping} --
As mentioned, when connecting to an anycasted service, traffic may -- in rare cases -- reach multiple anycast sites,
this is called \textit{anycast site flipping} and may lead to performance degradation of anycast.
One example is that it can break stateful TCP connections as state may only be kept at a single anycast site.

Wei et al.~measured the effects of anycast site flipping toward root DNS servers using DNS queries~\cite{wei2018does}.
They find that anycast is stable for 98\% of clients.
However, 1\% suffer from frequent flips between anycast sites.
For this 1\% they hypothesized that LB on-path to the anycast service caused the instability.
Using TCP-based DNS queries, they find that the measured instability rarely results in TCP timeouts.
Furthermore, studies by operators also suggest stateful services over anycast are functional (\ie, rarely affected by per-packet LB)~\cite{cachenetworks, linkedin}.
This suggests that most load balancers use per-flow or per-destination LB, as stateful services
keep these header fields static and therefore route consistently.

\vspace{0.1em}\noindent\textbf{Paris Traceroute} --
Traceroute is an Internet measurement tool that manipulates the TTL field
(hop-limit in IPv6) to trigger \texttt{ICMP Time Exceeded} replies at
intermediate hops toward a target.
By sending out packets with incrementing TTL,
with packet information encoded in header fields (including flow header fields),
traceroute is able to map the path to the target.
This path consists of a set of hops identified with their IP address and RTT\@.
However, in case of LB, traceroute can be misleading.
Depending on the LB decisions made, packets may take different paths to reach a target.
Furthermore, given that LB decisions are
made using the flow information of a packet, that are manipulated by
traceroute, the output may contain anomalous paths.
Paris traceroute solves this limitation~\cite{paris} by allowing the user to
define flow-identifiers and keeping them static.
Furthermore, it allows for manipulation of flow header fields to detect load balancers.

\vspace{0.1em}\noindent\textbf{Related Work} --
Various works in the literature looked at multi-path routing and LB\@.
In 2007, Augustin et al.~presented the Multi-path Detection Algorithm (MDA)
which builds on Paris traceroute to detect load balancers by
iteratively sending traceroutes with unique flow identifiers~\cite{mda}.
MDA is able to
locate multiple load balancers on a single path
and classifies LB behavior
by selectively changing header fields and evaluating which header
fields trigger LB decisions. 
The authors limited the work
in scale to a proof-of-concept due to high probing costs.
Their results showed per-flow LB is common, and per-packet LB is rare.

Ten years later, in 2017, Almeida et al.~performed a follow-up study for
LB on the IPv6 Internet using MDA~\cite{mdav6}.
Overall, their results showed that IPv6 LB was less widespread than IPv4.
Next, Vermeulen et al.~introduced MDA-Lite, providing a lower overhead implementation of MDA~\cite{mdalite}.
This work involved an IPv4 survey of multi-path routing,
showing load balancing topologies significantly increased in size and complexity.
Whilst MDA-Lite lowers overhead, it still requires a considerable number of traceroutes to detect LB,
especially when taking into consideration the increased size and complexity of multi-path topologies.

In 2019 and using the MDA algorithm as basis, Almeida et al.~designed the Multi-path
Classification Algorithm (MCA) and performed an extensive study of
LB for both IPv4 and IPv6~\cite{mca}. 
They find LB to be more prevalent than previous reports in their
2019 study: 74\% of IPv4 and 56\% of IPv6 routes traverse a load-balancer. 
Additionally, they find that LB is applied similarly across all transport protocols,
and that per-packet LB makes up only 0.1\% of load balancers.

A more recent study looked at intra-domain LB deployed at border routers that perform BGP-Multipath (BGP-M), to distribute traffic over multiple border links~\cite{bgpm}.
Using traceroute and Looking Glass servers they find large transit ASes and stub ASes widely deploy BGP-M to improve network performance.

\hspace{1cm}

Our methodology allows for detecting the effects of LB on an anycast deployment at Internet scale,
previously impossible due to high probing costs involved with traceroute in related work.
Furthermore, our methodology is able to detect LB in networks that perform tunneling, where traceroute has no visibility.
These methodologies can be combined;
our lightweight methodology detects clients experiencing site flipping at Internet scale,
and intensive traceroute-based techniques
pinpoint where LB decisions are being made.
Finally, the short measurement time and low burden of our approach allows operators to effortlessly observe the effectiveness of routing changes
when attempting to solve cases of site flipping.


\section{Methodology}\label{sec:methodology}

We use the TANGLED anycast testbed~\cite{tangled} to measure the effects of LB on anycast routing.
This testbed is currently hosted using Vultr, which offers 32 hosting locations spanning 6 continents and 20 
countries~\cite{vultr_locations}.
This provides us with 32 Vantage Points (VPs) from which to
announce a \texttt{/24} for measurement.
We capture responses to our probes at all VPs and by default send out probes from a single VP such that we ingress
ASes similarly between separate runs.
We target networks that appear on USC ISI's IPv4 hitlist~\cite{isihitlist}.
This list contains likely responsive IP addresses per \texttt{/24} prefix. 
For IPv6
we use the IPv6 hitlist from TU~M\"{u}nchen~\cite{v6hitlist}.
Thirdly, we also target \texttt{AAAA} record addresses for IPv6
and, as we consider DNS, name server IP address for IPv4.
These DNS records are provided by the
OpenINTEL project~\cite{openintel}. 

\vspace{0.2em}\noindent\textbf{Using Anycast to Infer Multi-Path Routing} --
As mentioned, this work is based on Verfploeter that infers the catchment of an anycast deployment~\cite{verfploeter}.
This approach involves sending probes to target networks, then based on where the replies end up it infers the catchment (\ie, mapping which target network routes to which anycast site).
Using Verfploeter as a basis, we provide a new methodology that
sends multiple probes to each target network, where each probe has a different flow header.

\begin{figure}
\includegraphics[width=0.5\textwidth]{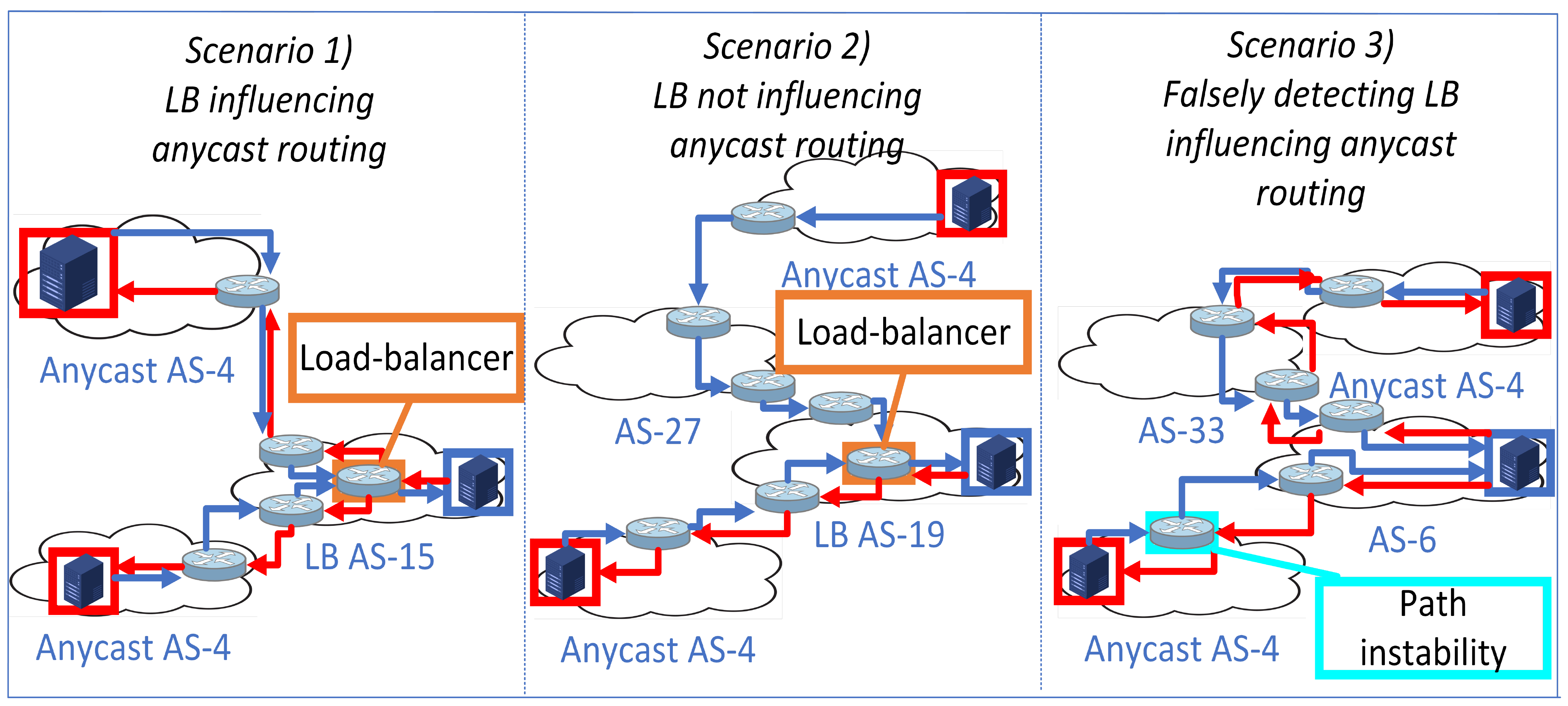}
\caption{Illustration of LB scenarios. Blue indicates how traffic from our anycast network would reach the probed unicast prefix, red indicates the possible paths that return traffic may take.}
\label{fig:scenarios}
\end{figure}

\vspace{0.1em}\noindent\textbf{Detecting Load-Balancing Effects on Anycast} ---
For LB to create anycast site flipping, the probed target must be in-between two or more of anycast VPs,
such that there are multiple routes available where LB decisions
causes traffic to be forwarded to distinct BGP paths that reach different anycast sites.
Based on this, there are three possibilities when it comes to
detecting anycast site flipping:

\emph{LB influencing anycast routing} --
When we probe a prefix in an AS that has multiple paths to our VPs,
we will observe LB behavior when that prefix gets routed
through a load-balancer.
Figure~\ref{fig:scenarios} shows three scenarios.
Scenario~1 has equal-cost paths to two of our anycast locations,
where based on LB decisions traffic may reach either anycast site.  
While the scenario shows LB inside a single AS, our approach works
regardless of where LB occurs (e.g., if an on-path AS performs LB).

\emph{LB not influencing anycast routing} --
The second option is that we probe an AS that performs LB, but only
has a single ``best'' path to one of our VPs. Then, regardless of the LB
decisions made we will receive replies at a single VP. The middle of
Figure~\ref{fig:scenarios} shows this scenario.

\emph{Falsely detecting LB influencing anycast routing} --
When a probed AS responds to multiple VPs due to behavior unrelated to LB (\eg, route flips), we will falsely attribute this to LB\@.
Scenario~3 shows this scenario, where traffic reaches two anycast sites due to path instability in the shortest path.

In theory, this methodology can also be used to detect ASes that perform LB on the Internet that fall under the second scenario.
We can create two equal-cost paths by manipulating BGP announcements
(using, \eg, BGP path prepending) to unveil ``hidden'' load balancers.
In our example we could prepend the BGP path
at our anycast location that directly peers with AS-19, giving the
load-balancer two equal-cost paths.
We leave exhaustive detection of prefixes behind load balancers using BGP path prepending as future work.

\vspace{0.2em}\noindent\textbf{Probing Methodology} ---
Our measurements focus on detecting the effects of LB on both IPv4 anycast and IPv6 anycast.
Also, we probe using the protocols ICMP, TCP and DNS\@.
This allows us to evaluate whether LB decisions are transport-protocol agnostic.

\begin{itemize}[itemsep=1pt,topsep=1pt,labelindent=0.25em,leftmargin=*]
    \item \textbf{ICMP} - we send out ICMP \texttt{ECHO} requests, for which we capture ICMP \texttt{ECHO} replies.
    \item \textbf{TCP} - we send out TCP \texttt{SYN|ACK} requests, for which we capture TCP \texttt{RST} replies.
    \item \textbf{UDP} - we send out DNS \texttt{A} queries, for which we capture responses.
    We discard ICMP \texttt{Unreachable} replies as they may originate from middleboxes.
\end{itemize}

We keep header fields static by default and carefully vary select header fields
for each measurement run to evaluate if the fields trigger LB behavior.
Each run uses 5 different values for the header field we vary,
as we observe diminishing returns beyond 5~values.
As a result, each target prefix receives 5 probes per measurement run.
We run ICMP, TCP and UDP measurements in which we vary the IP header.
For TCP and UDP we also perform
runs in which we vary the transport-layer ports.
We vary the IP header by probing
with multiple addresses within the anycasted prefix that we announce.
We do not measure LB using the \texttt{DSCP},  \texttt{Flow Label,} and \texttt{Traffic Class} fields
since middleboxes are known to overwrite these fields~\cite{mdav6, mca}.

Note that ICMP checksums will vary when ICMP payload changes.
This could trigger per-packet load-balancing.
As such, this
could create noise in runs in which we deliberately vary other fields.
We performed pre-measurement checks and found the effect to be negligible,
consistent with previous work that finds per-packet LB to be rare~\cite{mda,mca}.
For example, responses to multiple ICMP probes using non-varied IP headers arrived at
multiple of our VPs for only 1.4k prefixes, which is negligible ($<$\,0.02\%) compared to the
size of the hitlist and can be attributed to \eg, route flips.

\vspace{0.1em}\noindent\textbf{Anycast Traceroute} ---
We implemented a modified version of Paris traceroute that allows us to use an anycast source address.
For each probe reply we estimate the path length using the \texttt{TTL} value in the reply.
Next, we send out traceroute probes with incrementing \texttt{TTL} values up to the estimated path length.
To account for asymmetric paths, we add a margin of five additional hops in case the actual path is longer.
Since each traceroute probe reply may reach a different anycast site, we listen for \texttt{ICMP Time Exceeded} replies at all VPs\@.

We use this to estimate the AS at which load-balancing takes place, by evaluating at which hop we first see traceroute replies being received at multiple anycast sites.
However, this approach has limitations as this first hop may have a reverse path encountering a load-balancer in a different AS\@.

\vspace{0.2em}\noindent\textbf{Tooling} ---
We make our tooling publicly available as part of a larger toolchain that supports a wide range of anycast measurements~\cite{manycastr_tooling}.
The tooling is designed to be scalable for large anycast deployments, resource efficient for low burden at anycast sites, and containerized for easy deployment.
It supports IPv4 and IPv6 with underlying protocols ICMP, TCP, and UDP\@.
Furthermore, we allow for a large variety of configurations, including measurements using multiple combinations of port and address field values for outgoing probes.
The documentation of our toolchain clearly delineates how to measure the effects of LB on anycast routing.
We leave a detailed description of the tooling, designed for a large variety of anycast performance measurements, to future work.

\vspace{0.2em}\noindent\textbf{Limitations} ---
Our measurement setup has three limitations.
First of all the ISI hitlist is based on ICMP-responsiveness, meaning we will have fewer responsive targets for our TCP and UDP probes.
Especially for UDP, which is dependent on DNS services, responsiveness is low.
We include name server addresses provided by OpenINTEL to
create more responsiveness besides ICMP, but this still leaves us with a limited set of DNS-responsive targets.
Second, our methodology cannot reliably measure LB decisions that are dependent on traffic load,
though, we find that the low probing cost allows for repeated measurement over time
that can detect such cases.

Last, like Verfploeter we scan at a /24-granularity (/48 for IPv6), the smallest prefix size propagated by BGP\@.
However, as observed by Schomp et al.~ partitioning of the Internet is more fine-grained for some cases~\cite{partitioning},
where addresses belonging to a single /24-prefix are in topologically distinct locations causing traffic to route differently.
We limit this work to measure at /24 granularity as this assumption holds for most cases and to lower the measurement burden.
However, our methodology (and tooling) are agnostic to prefix size and operators can choose to measure at a finer granularity.
For example, whilst we probe at Internet-scale, operators can do tailored measurements for a much smaller number of destination prefixes at per-IP level granularity.

\vspace{0.2em}\noindent\textbf{Ethical Considerations} ---
Our research involves Internet-wide scans to identify LB behavior on a large scale, and it is performed under  IRB approval.
We have made efforts to conduct the measurements ethically and responsibly.
We maintain a low scanning rate - targeting 1 address in a single /24 - sending at most fifteen probes to a single target per hour.
In our TCP scan, we use a SYN-ACK based probing to avoid contributing to a possible state exhaustion on the target machines.
Additionally, probes are sent with a UTF-8 encoded URL in the payload, that refer to our experiments and abuse contacts, to account for possible opt-outs.


\begin{table}
\centering
\resizebox{.9\columnwidth}{!}{
\begin{tabular}{|c|r|r|c|}
        \hline
        \textbf{Protocol} & \textbf{Responsive} & \textbf{Affected by LB} & \textbf{\% affected by LB}\\
        \hline
        \hline
        ICMPv4 & 3,708,348 & 151,645 & 4.1\% \\
	\hline
	TCPv4 & 1,287,110 & 57,707 & 4.5\% \\
	\hline
	UDPv4 & 210,573 & 10,223 & 4.9\% \\
	\hline
	\hline
	IPv4 (total) & 3,820,285 & 167,573 & 4.4\% \\
	\hline
\end{tabular}
}
\vspace{0.1em}
\caption{The number of responsive /24-prefixes, and those that experience LB when varying the IP header,
listed by protocol.
\vspace{-2em}
}
\label{tab:overview}
\end{table}

\section{Results}
\label{sec:results}
We present our findings obtained using our anycast testbed.
This evaluation is the first study of load-balancing impacting anycast routing at Internet scale.
Though clients that experience anycast site flipping are deployment specific,
this section validates our methodology and provides insights to
the effects of LB on anycast routing.

\vspace{0.1em}\noindent\textbf{Prevalence of Load-Balancing} -- 
We probe for LB-behavior using ICMP, TCP, and UDP by varying the source
IP address, port values, or both.
For ICMP there are no port values,
for these prefixes we vary only the source IP address.
For IPv4 we have a hitlist of 5.6\,M prefixes.
To show the representativeness of our hitlist we show the number of
\texttt{/24}-prefixes responsive to each protocol in Table~\ref{tab:overview},
including those found to be affected by LB when varying the IP header.
These results show that we measured the effects of LB for 3,820,285 prefixes,
97.1\% of which respond to ICMP, 33.7\% respond to TCP, and 5.5\% respond to UDP.
Of those prefixes, we observe 167,573 (4.4\%) are routed to multiple anycast sites when varying
the IP header.
Using \textit{ip2location}~\cite{ip2location} we find that out of these 167,573 prefixes, 114,253 (68.2\%) are residential access networks.
As mentioned, when keeping header fields static we observe
a negligible number of prefixes respond to multiple anycast sites,
confirming prior work that finds per-packet LB is nearly non-existent~\cite{mda, mca}.
These results show that LB is prevalent,
and results in 4\% of the Internet reaching multiple anycast sites
based on LB decisions.

\begin{figure}[t]
\centering
\includegraphics[width=\columnwidth]{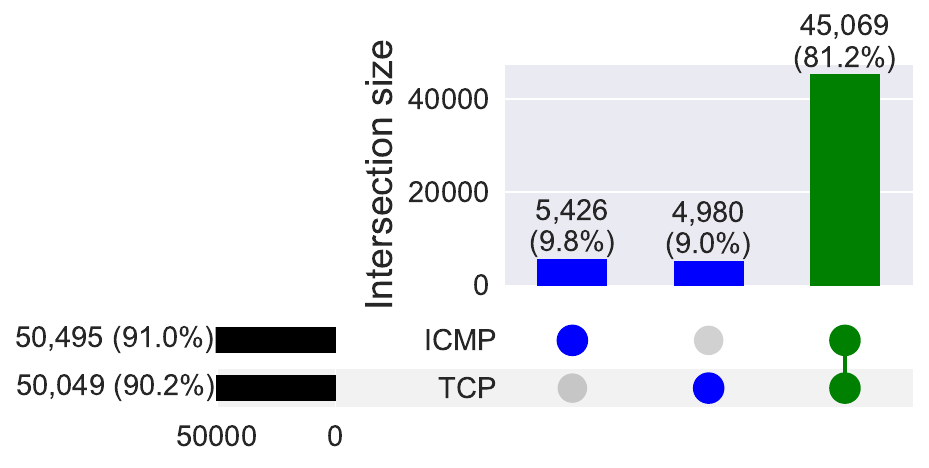}
\caption{LB by protocol for prefixes responsive to both ICMP and TCP.
}
\label{fig:protocol_lb}
\end{figure}

\vspace{0.1em}\noindent\textbf{Load-Balancing by Protocol} -- 
To assess whether LB is consistent among protocols,
we look at prefixes responsive to both ICMP and TCP\@.
In total, we have 1.2 million prefixes responsive to both protocols.
We exclude UDP from this analysis as we have few prefixes
responsive to DNS queries.
Figure~\ref{fig:protocol_lb} shows the results using an \textit{UpSet} plot.
On the bottom-left are two counts for prefixes found to be
affected by LB for the ICMP and TCP runs when varying the IP header.

In these two runs combined, we detected 55,475 prefixes 
that exhibit LB behavior for either protocol.
We detected 50,495 (91.0\%) display LB behavior in the ICMP
run and 50,049 (90.2\%) for the TCP run.
The bars in the figure show counts for the complement of the intersections of prefixes found in the
two measurement runs.
We color intersections by degree:
\texttt{blue} for prefixes found in only a single measurement run,
\texttt{green} for prefixes found in both measurement runs.
Bar~1 shows the 5,426 prefixes found exclusively in the ICMP run (9.8\% of the total).
Bar~2 shows 4,980 prefixes exclusively in the TCP run (9.0\%).
Bar~3 accounts for the 45,069 prefixes found in both runs (81.2\%).
\noindent
We believe the non-intersecting prefixes (bars 1 and 2) are due to noise between runs rather than protocol-specific LB\@.
Possible reasons for this noise are:
\begin{itemize}[itemsep=1pt,topsep=1pt,labelindent=0.25em,leftmargin=*]
	\item \textbf{Churn} - The probed target may not have been responsive for all measurement runs.
	\item \textbf{Route flips} - Prefixes may route differently between runs due to route flips,where they route through an AS performing LB in certain measurement runs, but get routed through an AS that does not perform LB in other measurement runs.
	\item \textbf{Load dependent} - Some LB behavior may only trigger when load is present, which is dynamic over time.
	\item \textbf{Hashing changes over time} - Load-balancers may change their hashing function over time~\cite{dminer}.
\end{itemize}

The majority of LB is detectable in both measurement runs (bar 3),
indicating that LB is consistent among protocols, as backed by previous work~\cite{mca}.
In contrast, we observe inconsistency in Table~\ref{fig:protocol_lb},
we suspect this is caused by bias as to where TCP and DNS responsive prefixes are located.

\begin{figure}[t]
\centering
\includegraphics[width=\columnwidth]{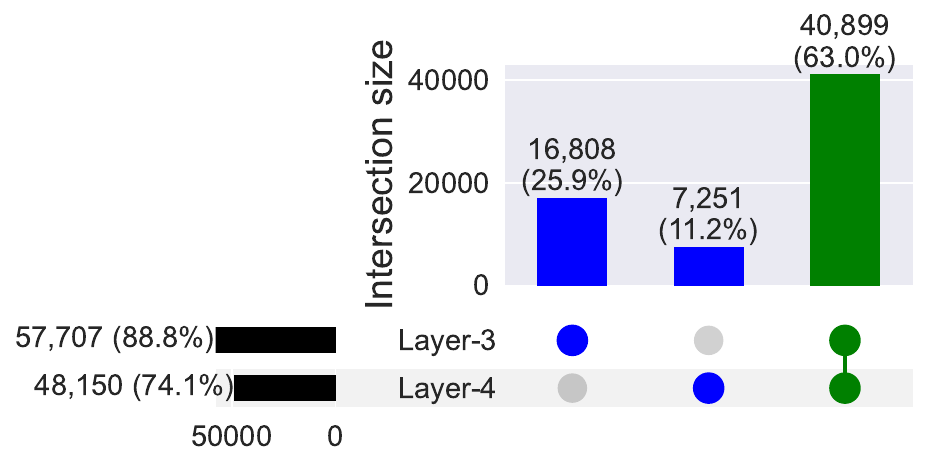}
\caption{LB detected when varying the IP-header (layer 3), the TCP-header (layer 4), or both (layer 3 \& 4) for TCPv4.
}
\label{fig:layer_lb}
\end{figure}

\vspace{0.1em}\noindent\textbf{Load-Balancing by Network Layer} -- 
As mentioned, router vendors have different configurations for LB hashing functions.
As shown in Table~\ref{tab:vendors}, most default configurations calculate the hash using the entire flow 5-tuple (layer 3 \& 4) and some use only the IP header (layer 3).
To measure LB configurations deployed by operators we perform TCP runs varying either the IP source address (layer 3) or TCP port fields (layer 4),
the results are shown in Figure~\ref{fig:layer_lb}.

Overall, we observe most LB behavior when varying the IP-header (57,707)
and slightly less when varying the TCP-header (48,150).
This is unsurprising as all vendor configurations use the IP-header,
whereas only flow-based LB configurations use the TCP-header.
We find a considerable number of LB detected using only layer 3 (16,808).
Surprisingly, we also observe 7,251 (0.6\% of TCP responsive prefixes) cases of layer 4 only LB, inconsistent
with vendor default configurations.
This behavior was also observed in previous work~\cite{mca}, but not as prevalent.
We suspect this is an effect of noise, as explained previously.
Finally, we find most load balancers (40,889) hash using both layers, \ie, per-flow LB\@.

\begin{table}
\resizebox{\columnwidth}{!}{
\begin{tabular}{|ll||r|r|r|}
        \hline
        \textbf{AS} & \textbf{AS Name} & \textbf{LB /24s} & \textbf{Responsive} & \textbf{Ratio LB}\\
        \hline
        \hline
 	7303 & Telecom Argentina & 17,837 & 19,037 & 0.937 \\
	\hline
 	3352 & Telef\'{o}nica de Espa\~{n}a & 15,216 & 23,197 & 0.656 \\
	\hline
 	8075 & Microsoft & 13,121 & 64,869 & 0.202 \\
	\hline
 	8708 & Digi Communications & 5,432 & 5,513 & 0.985 \\
	\hline
	3462 & Chunghwa Telecom & 4,713 & 9,979 & 0.472 \\
	\hline
	36947 & Telecom Algeria & 4,357 & 5,293 & 0.823 \\
	\hline
 	5769 & Videotron & 4,156 & 8,663 & 0.480\\
	\hline
 	16509 & Amazon & 4,128 & 87,184 & 0.047 \\
	\hline
	9198 & JSC Kazakhtelecom & 3,936 & 4,198 & 0.983 \\
	\hline
	6730 & Sunrise & 3,436 & 5,145 & 0.668 \\
	\hline
\end{tabular}
}
\vspace{0.1em}
\caption{10 largest ASes affected by LB, the number of responsive prefixes,
and the ratio of prefixes affected by LB (for ICMP).
\vspace{-2em}
}
\label{tab:largest_icmp}
\end{table}

\vspace{0.1em}\noindent\textbf{Load-Balancing by Autonomous System} --
For ICMPv4 we observe LB behavior in 4,270 ASes.
Table~\ref{tab:largest_icmp} shows the 10 largest ASes by number of LB /24-prefixes detected.
This table shows the number of prefixes affected by LB, the total number of ICMP-responsive prefixes, and the ratio.
We find the majority of prefixes that experience anycast site flipping originate from eyeball networks,
where most prefixes experience site flipping for some
(Telecom Argentina, Digi Communications, JSC),
and only a portion of prefixes for others
(Telef\'{o}nica de Espa\~{n}a, Chungwha, Videotron, Sunrise).
We believe the latter, \ie, partially affected by LB, to be caused by topological differences inside the AS\@.
Finally, we see two hypergiants (Microsoft, Amazon) partially affected by LB\@.

\begin{figure}[t]
  \centering
\includegraphics[width=\columnwidth]{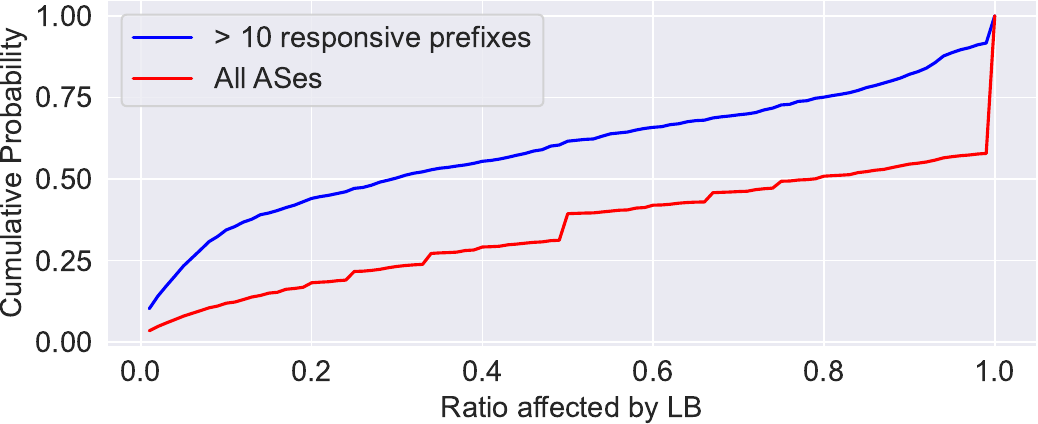}
\caption{Ratio of prefixes affected by LB and responsive prefixes, by AS (for all ASes and ASes with more than 10 responsive prefixes).
}
\label{fig:ratio}
\end{figure}

To give insight into the percentage of prefixes affected by LB, we show a CDF in Figure~\ref{fig:ratio}.
We plot all ASes (red), and ASes for which we have more than 10 responsive prefixes (blue).
Looking at the filtered ASes (blue), we find roughly a third of ASes to have less than 10\% of prefixes affected by LB\@.
we suspect these to be cases of \eg, route flips rather than LB\@.
Further, we observe 40\% of ASes to have more than half of their prefixes affected by LB,
and 20\% of ASes have the vast majority affected by LB\@.
Evaluating the results using all ASes (red), we see at the top right of the graph that more than 40\% of ASes
are entirely affected by LB\@.
These are for the vast majority small ASes where we observe anycast site flipping for all responsive prefixes.
Overall these results show that large ASes are often partially affected by LB,
indicating different topologies,
and we observe a large number of small ASes that are affected entirely.

\begin{table}
\resizebox{\columnwidth}{!}{
\begin{tabular}{|ll||r|r|r|r|r|}
        \hline
        \textbf{AS} & \textbf{AS Name} &  \multicolumn{1}{c|}{\textbf{IP}} &  \multicolumn{1}{c|}{\textbf{SRC}} & \multicolumn{1}{c|}{\textbf{DST}} &  \multicolumn{1}{c|}{\textbf{SRC+DST}}&  \multicolumn{1}{c|}{\textbf{All}}\\
        \hline
        \hline
 	7303 & Telecom Argentina & 8,828 & 9,188 & 9,052 & 8,401 & 9,084 \\
	\hline
	3462 & Chunghwa Telecom & 4,992 & 4,148 & 4,905 & 4,843 & 5,016 \\
	\hline
 	6713 & Maroc Telecom & 3,889 & 4,716 & 4,633 & 4,667 & 4,644 \\
	\hline
 	1267 & Wind Tre & 2,002 & 2,010 & 1,979 & 2,002 & 1,993 \\
	\hline
	36947 & Telecom Algeria & 1,312 & 1,378 & 1,356 & 1,244 & 1,391 \\
	\hline
	\hline
	25820 & IT7 Networks & 585 & 0 & 0 & 0 & 586 \\
	\hline
	18779 & EGI Hosting & 579 & 121 & 187 & 188 & 585 \\
	\hline
	9676 & Savecom & 322 & 0 & 0 & 0 & 327 \\
	\hline
	11830 & Costarricense Telecom & 310 & 131 & 133 & 129 & 314 \\
	\hline
	7018 & AT\&T & 293 & 120 & 120 & 120 & 263 \\
	\hline
\end{tabular}
}
\vspace{0.1em}
\caption{5 largest ASes affected by LB, followed with the 5 largest ASes showing bias toward layer-3 only LB (for TCP).
\vspace{-2em}
}
\label{tab:largest_tcp}
\end{table}

Next, we look at the largest ASes affected by LB for TCP,
where we can differentiate between layer 3 and layer 4 LB\@.
Table~\ref{tab:largest_tcp} shows the 5 largest ASes by number of LB /24-prefixes for TCP,
followed with the 5 largest ASes that have a clear bias toward layer 3 only LB\@.
For TCP we find a different ranking of ASes compared to ICMP,
we attribute this to difference in responsiveness.
The table contains the number of LB /24s found for separate measurement runs varying either the
source address (\texttt{IP}), source port (\texttt{SRC}), destination port (\texttt{DST}),
both port fields (\texttt{SRC+DST}), and source address and both port fields (\texttt{All}).
We observe that most ASes perform LB consistently for all variations of header fields,
and use both layer 3 \& 4 headers (\ie, per-flow).
Finally, we observe 35 ASes only visible when varying layer 4 fields,
all of which have few prefixes showing LB behavior.

\vspace{0.1em}\noindent\textbf{Measurement Consistency} -- 
As mentioned, our methodology experiences noise (\eg, prefixes not always seen to be affected by LB).
To quantify this noise we run 10 back-to-back ICMPv4 measurements.
These runs observe anycast site flipping ranging from 154k to 158k prefixes.
For 87.2\% of prefixes the site flipping is observed at all 10 measurement runs,
for 10.4\% of prefixes it is seen in 3--9 measurement runs,
for 0.8\% of prefixes it is detected in 2 measurement runs,
and for 1.6\% of prefixes it is observed once.
Prefixes seen in all measurement runs are clear cases of LB,
those found in 3--9 measurement runs we suspect to be cases of load-dependent LB,
and the remaining prefixes seen at only one or two measurement runs we attribute to route flips.

For the first measurement we find 153,870 prefixes,
of which 152,130 are visible again in the second measurement run 20 minutes later.
However, comparing the first to the last measurement (time difference of 3 hours and 5 minutes)
we observe an intersection of 149,324 prefixes.
We suspect that this decrease in intersecting prefixes
is due to changes in traffic load and route flips, between LB paths and non-LB paths, that are more likely to occur over time.

\vspace{0.1em}\noindent\textbf{Long-lived Anycast Site Flipping} --
To determine whether measured anycast site flipping caused by LB is persistent for long periods of time,
we compare data with a 4-month interval.
We look at prefixes found to route toward multiple anycast sites when varying
the IP header of ICMPv4 probes, we use data from 15 February 2024 and 21 June 2024.
For the first, we have 153,734, and for the second we have
151,645 prefixes.
Taking the intersection, we find 89,447 prefixes are affected by LB
for both measurement runs.
This suggests most anycast site flipping is long-lived.
However, a significant part does appear to be short-lived that would require constant monitoring from operators.
We leave an explanation of such short-lived instability to future work.

\vspace{0.1em}\noindent\textbf{IPv6} -- 
We perform IPv6 measurements on a \texttt{/48} prefix granularity,
as this is the smallest (globally) routable prefix size.
Table~\ref{tab:overview_v6} shows the responsiveness and number of prefixes affected by LB for ICMPv6, TCPv6, UDPv6.

\begin{table}
\centering
\resizebox{0.9\columnwidth}{!}{
\begin{tabular}{|c|r|r|c|}
        \hline
        \textbf{Protocol} & \textbf{Responsive} & \textbf{Affected by LB} & \textbf{\% affected by LB}\\
        \hline
        \hline
        ICMPv6 & 531,663 & 20,193 & 3.8\% \\
	\hline
	TCPv6 & 150,730 & 3,738 & 2.5\% \\
	\hline
	UDPv6 & 11,303 & 495 & 4.4\% \\
	\hline
	\hline
	IPv6 (total) & 534,611 & 20,932 & 3.9\% \\
	\hline
\end{tabular}
}
\vspace{0.1em}
\caption{The number of responsive /48-prefixes, and those that experience LB when varying the IP header,
listed by protocol.
\vspace{-2em}
}
\label{tab:overview_v6}
\end{table}

Overall, we find there are fewer prefixes experiencing anycast site flipping due to LB compared to IPv4,
confirming previous work that finds LB is less prevalent for IPv6~\cite{mca}.
For TCP and UDP our hitlist lacks responsive targets for meaningful analysis.
For ICMP, we observe LB for 20,193 \texttt{/48s} located in 854 ASes.
The vast majority, 11,685 (57.9\%), of prefixes affected by LB are announced by Cloudflare (AS\,13335) which we discuss later.

\vspace{0.1em}\noindent\textbf{Validation using Paris Traceroute} -- 
We validate our results using Paris traceroute,
the current state-of-the-art for detecting LB\@.
As mentioned, we implemented a modified version of Paris traceroute that works with an anycast source address.
However, performing Paris traceroute from our anycast testbed measures LB on-path from deployment to client,
whereas our methodology detects LB from client to deployment.
Therefore, we use the RIPE Atlas platform~\cite{staff2015ripe}
that provides more than 11k distributed VPs for
Internet measurement, including support for Paris traceroute.
This allows us to perform Paris traceroute from the client to the deployment to validate our methodology.

First, we detect RIPE Atlas VPs that experience anycast site flipping using our methodology by varying the IP header.
In total we find ~7k RIPE Atlas VPs that are ICMP-responsive,
of which 306 VPs (located in 152 distinct ASes) reach multiple anycast sites when varying the IP header.
Next, we perform Paris traceroutes from all RIPE VPs
toward multiple addresses in the \texttt{/24} we anycast, using the same IP header variations.
This measurement campaign lasted several days due to RIPE Atlas constraints.
To identify the anycast site reached in the traceroute logs,
we mimic each site to be routed behind an additional fake host using a unique hop address.
In total, 9,613 RIPE VPs were available for traceroute measurement
of which 9,213 revealed the reached sites in their traceroute logs.
For 240 out of 306 RIPE VPs where our methodology detects anycast site flipping,
the LB behavior is validated with traceroute.
The missing 66 VPs were either unavailable for traceroute measurements
or failed to reveal LB behavior due to traceroute limitations (e.g., networks blocking \texttt{ICMP Time Exceeded} replies).
For the validated VPs, we observe a consistent mapping of IP-header to anycast site reached,
\ie, load balancers consistently split traffic toward the same sites
and each IP-header variation is observed to reach the same anycast site using both methodologies.

Investigating RIPE Atlas traceroutes from all VPs we find an additional 180 RIPE VPs that
experience anycast site flipping that are missed by our methodology.
We find that 155 of these VPs are unresponsive to ICMP
(\ie, undetectable with our methodology),
the remaining 25 we suspect to be caused by the time difference between measurements, allowing for load-dependent LB and route flips to occur.

\vspace{0.1em}\noindent\textbf{Determining where Load-Balancing takes place} -- 
To determine where LB decisions creating anycast site flipping take place,
we look at the 240 RIPE Atlas VPs observed to experience site flippin using both methodologies.
For this we use traceroute results from RIPE VPs toward TANGLED (our anycast deployment)
and traceroute results from TANGLED toward RIPE VPs.
We perform the latter using our implementation of Paris traceroute
with an anycast source address,
that detects hops behind load balancers based on \texttt{ICMP Time Exceeded} replies reaching multiple anycast sites.

\begin{figure}[t]
\centering
\includegraphics[width=0.9\columnwidth]{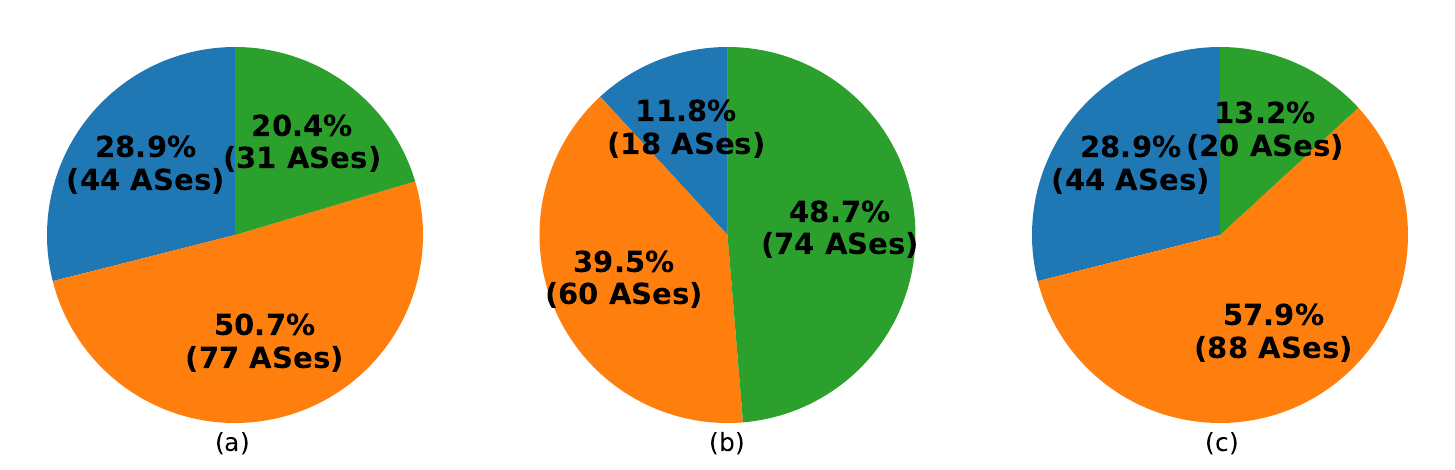}
\caption{
Where LB takes place for RIPE Atlas VP ASes experiencing anycast site flipping: \textbf{blue} -- at RIPE Atlas VP AS, \textbf{orange} -- at on-path AS, \textbf{green} -- unknown location.
Measured using Paris traceroute from: (a) RIPE Atlas VPs, (b) anycast deployment, (c) combining both directions.
}
\label{fig:lb-location}
\end{figure}

We differentiate between three cases for where LB, causing anycast site flipping, takes place;
\texttt{Home AS} - the load-balancer is in the same AS as the client,
\texttt{On-path AS} - the load-balancer is in an AS between the client and anycast deployment,
\texttt{Unknown} - it is unknown where the load-balancer is (\eg, due to networks using MPLS).
Figure~\ref{fig:lb-location} shows where we detect LB to take place,
when performing from the client toward the anycast deployment (a),
when performing from the deployment toward the client (b),
and when combining traceroute data from both directions (c).
Comparing the first two cases (a, b), we observe visibility
is considerably higher when performing traceroute from the client
where for only 20.4\% of affected ASes it remains unknown where LB takes places,
compared to 48.7\% when performing traceroute from the deployment.
We attribute the latter to routing asymmetry, in which the path between the deployment and the client differs from the path between the client and the deployment where LB takes place.
Furthermore, we observe this lack in visibility to be especially present
for cases where LB takes place at the home AS, 
which we attribute to networks only visible with traceroute from within and upstream routers that block \texttt{ICMP Time Exceeded} replies.

Combining traceroute data from both sources (c), we find that for 88 ASes (57.9\%)
that experience anycast site flipping,
LB occurs in on-path ASes.
For 44 ASes (28.9\%) the LB originates from within.
For the 20 remaining ASes (13.2\%) the origin of LB is unknown due to traceroute limitations.
These results show that most LB, creating anycast site flipping, takes place in on-path ASes.
Further, we find that traceroute has significant limitations when detecting LB,
especially when performed from the anycast deployment where the LB is invisible for 48.7\% of affected ASes.

\vspace{0.1em}\noindent\textbf{Eyeball Networks} --
As shown in Table~\ref{tab:largest_icmp}, a large number of our results originate from
telecom providers.
Additionally, using \textit{ip2location}~\cite{ip2location} we find 68.2\%
of IPv4 prefixes where we detect anycast site flipping are residential.
Private communication with an operator of one of these eyeball
networks confirmed that the behavior we observed is consistent with
their routing LB implementation.
More specifically, they find that OSPF based ECMP (at CMTS level)
results in traffic reaching different transit providers,
this LB implementation is supported by most router vendors (\eg, HPE~\cite{ospf}).

For our TCPv4 results the majority of prefixes also originate from
eyeball networks , as shown in Table~\ref{tab:largest_tcp}.
For some ASes, we observe LB behavior only visible when varying port fields though they are few and have few prefixes affected by LB\@.
This is consistent with results from previous work~\cite{mca}, that saw 2\% of load-balancers hashing using only layer 4 fields.

\begin{figure}[t]
\includegraphics[width=\columnwidth]{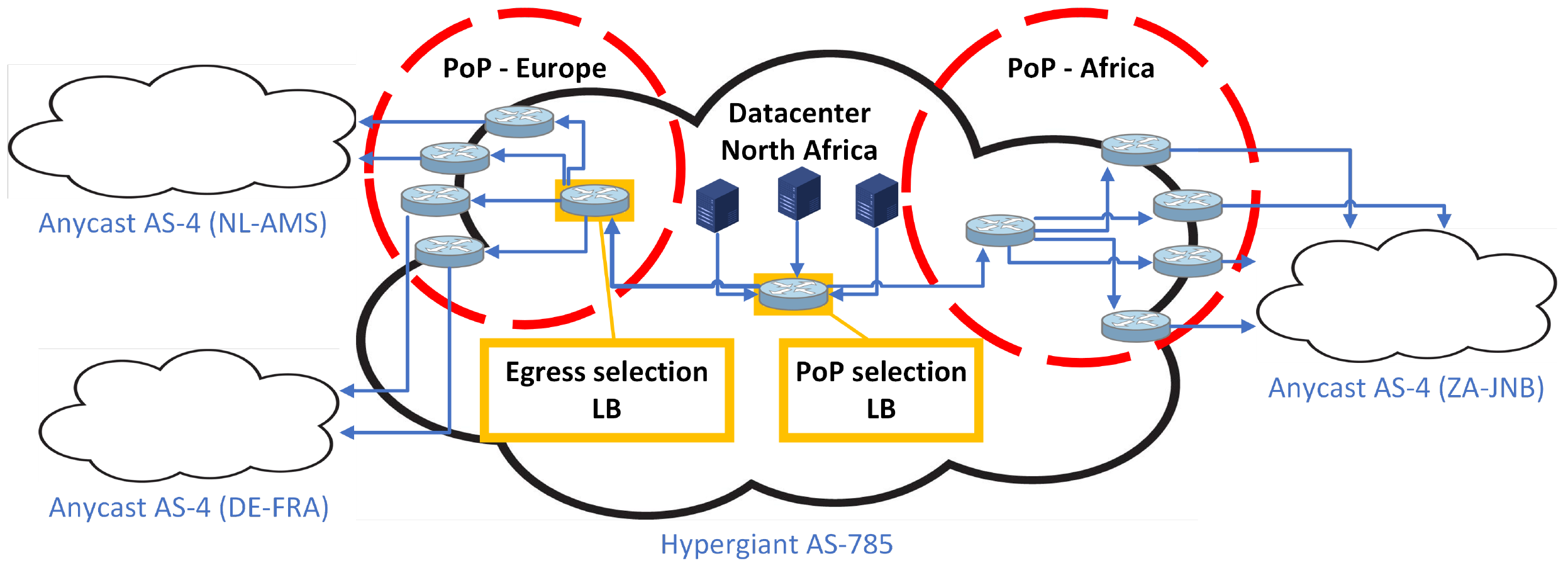}
\caption{LB happening at the Point of Presence (PoP) and/or egress selection within a hypergiant's network.}
\label{fig:hypergiant_simple}
\end{figure}

\vspace{0.1em}\noindent\textbf{Hypergiants} --
Several so called ``hypergiants'' originate 24.9\% of IPv4 prefixes where we detect LB\@.
Such hypergiants have their own private backbone infrastructure consisting
of their own private links to interconnect Points of Presence (PoPs).
Each PoP connects to the global Internet using a set of egress ports
(which can range from tens to hundreds at a single PoP)~\cite{facebook}.
Such sophisticated edge networks mean that there are multiple places where
load-balancing decisions can be made.
To give a simplified example, we provide an illustration in Figure~\ref{fig:hypergiant_simple}.
This is also reflected in our results, where we observe that replies destined for a
single unicast address -- within a hypergiant's network -- are routed back to
geographically diverse anycast VPs (sometimes even VPs located on different continents).
The largest hypergiant in our results is
Microsoft's AS\,8075 (13k prefixes in our ICMPv4 measurements).
To further investigate our observations for Microsoft prefixes
we created VMs on Azure to measure connectivity to our anycast deployment from within their infrastructure.
When creating VMs they provide several load-balancing options~\cite{microsoftLB}, and we find the options supporting cross-region LB
to route toward distinct anycast sites when varying the IP header.
These findings show that hypergiants, such as Microsoft, offer a wide variety of routing options
that can lead to anycast site flipping.

Next, we reached out to a large CDN network operator,
where we observed reply traffic reaching VPs in different continents.
We confirmed through spot-checks that the identified LB behavior was due to BGP-M path selection from a PoP to one of the CDN's upstreams.
Finally, we reached out to operators at Cloudflare,
by far the most prevalent AS in our ICMPv6 result consisting of 12k prefixes.
From these talks we learned that the prefixes in our results are unicasted
within Cloudflare's network (\ie, our probes end up at a single location),
however the BGP announcements for these prefixes are propagated at multiple Cloudflare PoPs.
Our anycast Paris traceroute results confirm this, as we see our probes ingressing at distinct locations.
However, due to limitations in our methodology we cannot observe the return path.
From these talks, we further learned that Cloudflare TE implementation
egresses near the client using their geofeed~\cite{geofeed}.
However due to the anycasted nature of our IP,
this geofeed wrongly believes our IP to be located at a single location.
Therefore, reply traffic for our prefix will egress at the same PoP
regardless of the location of the probed address within Cloudflare's infrastructure.
From this PoP, Cloudflare traffic reaches multiple sites due to
load balancers located in on-path ASes.

\vspace{0.1em}\noindent\textbf{Major Transit Providers} --
Like hypergiants that have multiple PoPs, major transit providers interconnect the Internet at geographically distributed Internet Exchange Points (IXPs).
We observe major transit providers -- in some cases -- prefer routes toward our anycast IP using IXPs in distant locations,
when a short path toward a nearby site is available.
In total, 9 major transit providers show this behaviour in our dataset.
An illustrative example is Hurricane Electric, where using their Looking Glass service we observe routers in Europe prefer paths through an IXP in India for our anycasted IPv6 prefix.
Ultimately, we observe this results in a large number of IPv6 prefixes being load-balanced between two sites in India.
It is unclear whether this load-balancing occurs at the transit provider or afterward.

\vspace{0.2em}\noindent\textbf{Multiple-Client Probing} --
In \S\ref{sec:methodology} we explained that ICMP probes with
non-varied IP header fields from a single VP prompted 1.4k prefixes ($<$\,0.02\%) to respond to multiple VPs.
To evaluate if probing from multiple locations instead of a single VP impacts LB decisions,
we perform a measurement where we probe using four VPs on different continents.
This reveals 21k prefixes whose replies are received at multiple VPs.
Since we effectively perform a \manycast~census~\cite{manycast}, we also capture anycast behavior.
For this reason we filter out anycasted prefixes using an anycast census~\cite{laces}.
This leaves us with 9.6k prefixes that are not anycast.
We confirm these prefixes are unicast with iGreedy's latency-based anycast detection methodology~\cite{igreedy} using RTTs captured from CAIDA's Ark platform~\cite{ark}.
Of these unicast prefixes, 5.9k originate from Microsoft's AS\,8075.
Using our anycast Paris traceroute, we see that these prefixes are announced at multiple Microsoft PoPs
and that each of our anycast VPs are ingressing at different Microsoft PoPs.
We believe these prefixes are ending up at multiple sites due to
symmetrical ingress/egress routing policies,
that cache the observed ingress path for a packet and forward the egress traffic through the same path.

\begin{figure}[t]
\includegraphics[width=0.5\textwidth]{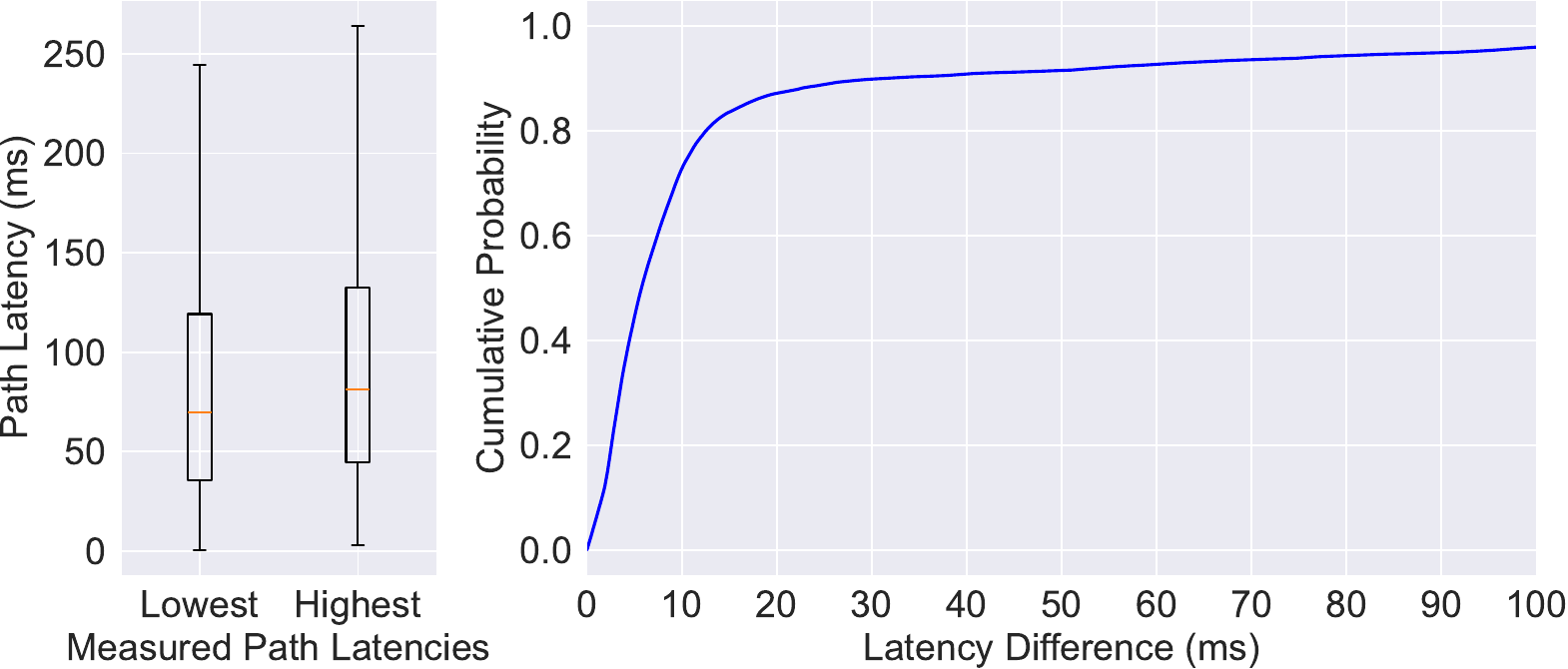}
\caption{Boxplot (left) for the lowest and highest average latency paths,
and a CDF (right) for the latency difference between the two.
}
\label{fig:latencies}
\end{figure}

\vspace{0.2em}\noindent\textbf{Multi-Path Latencies} --
Our methodology finds networks that route to multiple anycast sites,
each of these routes have different path latencies.
If an operator uses our methodology to detect these networks,
they can implement traffic engineering, \eg, using selective BGP announcements,
to direct clients toward the shortest (in terms of latency) path.

Figure~\ref{fig:latencies} shows a boxplot and a CDF\@.
First, the boxplot contains the average latencies observed for the shortest path (left)
and the longest path (right).
This latency is the time difference between transmission time at the sender (site in Amsterdam) that probes the client,
and the receiving time of the anycast site that receives the client probe reply.
The average latency for the shortest path is 88.3 ms
and 103.2 ms for the longest path, giving a difference of 14.9 ms.
In normal circumstances, where the client communicates
to and from the same site this translates to a difference of 29.8 ms in RTT since the inflated path is used in both directions.
The latency difference is dependent on the anycast deployment,
for reference, our anycast deployment has an average latency of 60.0 ms
toward targets on the IPv4 hitlist.

Next, the CDF shows the distribution of the latency difference 
for prefixes with multiple load-balanced paths toward our deployment.
We find that for 75\% of prefixes the latency difference is at most 10 ms.
However, for the remaining 25\% the CDF shows a long tail
where 4\% of prefixes have a latency difference of more than 100 ms.
This long tail contains prefixes whose traffic is load-balanced between
sites in different continents.
These results show that operators can significantly decrease latency
for clients experiencing anycast site flipping when directing them to the best path.
We leave a more in-depth analysis as to when LB causes site flipping with significant latency differences to future work.


\section{Conclusions \& Future Work}
\label{sec:conclusion}

We introduced a novel methodology that detects networks
experiencing anycast site flipping caused by load-balancing.
We publicly share our tooling~\cite{manycastr_tooling} that detects these cases at Internet scale with a low probing cost and short measurement time,
impossible with previous tooling that depends on traceroute.
We find that 96\% of probed Internet prefixes consistently route to a single anycast site.
However, for the remaining 4\% we detect site flipping where based on load-balancing decisions traffic reaches multiple anycast sites.
We find that, when keeping flow headers static, load balancers will consistently route clients to the same anycast site.
These results show that stateful services can be offered using anycast, and as such can benefit from the resilience that anycast offers.

However, services that require multiple stateful sessions will be harmed by LB
as individual sessions may reach different anycast sites.
Furthermore, route flips may yet harm long-lasting stateful connections,
we leave a study on the stability of anycast routing for long-lasting connections to future work.

Whilst our initial objective was to detect the effects of load balancers on anycast routing,
our results show that different types of traffic engineering implementations result in anycast site flipping.
These implementations range from load balancers to PoP- and egress-selection mechanisms inside intercontinental hypergiant networks.

We believe operators benefit from knowing when clients connect to
multiple anycast sites for, \eg, catchment analysis and can
solve these cases using, \eg, BGP path prepending and path poisoning.
However, such changes in routing may shift the problem to different networks,
requiring monitoring of effectiveness that our tool makes possible.
Finally, we find that the RTT these clients experience can be significantly improved
when directing them to the nearest site, potentially lowering RTT by hundreds of milliseconds for some clients.

\section*{Acknowledgments}
This work was supported by
INTERSCT (NWO grant NWA.1160.18.301).


Any views and opinions expressed in this work are those of the
authors and do not necessarily reflect those of
the Dutch research funding agency NWO\@.
NWO cannot be held responsible for them.

\bibliographystyle{IEEEtran}
\balance
\bibliography{ecmpcast}

\end{document}